\documentclass[11pt,onecolumn,amssymb,nofootinbib]{revtex4}
\usepackage{amsmath, amsthm, amscd, amssymb}
\usepackage{bm}
\usepackage{bbm}

\begin{document}

\title{\bf Free field structure of the model with a spin-$\frac{3}{2}$ Rarita-Schwinger field
directly coupled to a spin-$\frac{1}{2}$ field }

\author{Stephen L. Adler}
\email{adler@ias.edu} \affiliation{Institute for Advanced Study,
Einstein Drive, Princeton, NJ 08540, USA.}

\begin{abstract}
In earlier work we introduced an abelianized gauge field model in which a Rarita-Schwinger field is directly coupled to a spin-$\frac{1}{2}$ field, and showed that this model admits a perturbative expansion in the gauge field coupling.  As a preliminary to further study of the coupled model, in this paper we
present a detailed  analysis of the free field structure that obtains when the dimensionless gauge coupling is set to zero, but the dimension one coupling of the spin-$\frac{3}{2}$ and spin-$\frac{1}{2}$ fields remains  nonzero.
\end{abstract}

\maketitle
\section{Introduction}
This paper is a continuation of our study of whether Rarita-Schwinger fields can be consistently gauged.
In an earlier paper \cite{adler1} we introduced a model in which a massless spin-$\frac{3}{2}$ field is coupled to a spin-$\frac{1}{2}$  field with a coupling parameter with dimensions of mass, and showed that this model, unlike the standard Rarita-Schwinger theory, can be gauged without introducing singularities at zero gauge field. A detailed discussion of motivations, and earlier references, is given in the introductory section of \cite{adler1} and will not be repeated here; suffice it to say that the question of the gauging of Rarita-Schwinger fields is of long standing, and is of interest both from the vantage point of the possible use of a gauged spin-$\frac{3}{2}$ sector in unified models, and the theory of gauge anomalies of spin-$\frac{3}{2}$ fields. Despite admitting a perturbative expansion, and permitting calculation of the gauge anomaly,  the
coupled model was shown in \cite{adler1} to have a number of non-standard features: (1)  The zero external field plane wave solutions are not all eigenvectors of the wave matrix, but rather, some of the plane wave solutions are only Jordan canonical form eigenvectors.   (2) There are tachyonic propagation modes in the presence of external fields.   (3) The Dirac brackets are non-positive.
We suggested in \cite{adler1} that some of these problems may be cured when dynamical symmetry breaking is
taken into account.  For example, the tachyonic modes could be a signal that the model is unstable against
symmetry breaking  giving the fields of the model
masses and altering the propagating modes in the presence of gauge fields.  And such symmetry breaking could lead to a Lee-Wick \cite{leewick} resolution of the negative metric problem, for example by making the negative metric sector modes unstable against decay into positive metric sector modes.

To assess whether these suggestions are viable will require a further detailed study of the dynamics of the
interacting theory.  But a first essential step is to understand the free field, zero gauge coupling  structure of the model in detail, and that is the purpose of this paper.  In Section II, we give a systematic mode analysis in momentum space, determine the equal time anticommutator algebra of the creation--annihilation operators, and isolate the  indefinite metric sector.  In Section III, we study the left-chiral propagator,  its relation to time-ordered products of fields, and its construction from the action of creation and annihilation operators on the vacuum, and
the Hamiltonian time evolution of these operators.  Brief conclusions are given in Section IV.

\section{Mode analysis}
\subsection{Lagrangian, field equations, and anticommutators}
The Lagrangian density and Euler-Lagrange equations of the model in manifestly covariant form are
\begin{align}\label{covL}
{\cal L}=& -\overline{\psi}_{\mu} \gamma^{\mu\nu\rho} \partial_{\nu}\psi_{\rho}
-\overline{\lambda}\gamma^{\nu}\partial_{\nu}\lambda+m(\overline{\lambda}\gamma^{\nu}
\psi_{\nu}-\overline{\psi}_{\nu}\gamma^{\nu}\lambda)~~~,\cr
0=&\gamma^{\mu\nu\rho}\partial_{\nu}\psi_{\rho}+m\gamma^{\mu}\lambda~~~,\cr
0=&\gamma^{\nu}\partial_{\nu}\lambda-m\gamma^{\nu}\psi_{\nu}~~~,\cr
\end{align}
with $\psi_\mu$ the spin-$\frac{3}{2}$ Rarita-Schwinger field and $\lambda$ the spin-$\frac{1}{2}$ field.
From taking $\partial_{\mu}$ of the second line we find that $\lambda$ obeys a free
massless Dirac equation,
\begin{equation}\label{lameq}
0=\gamma^{\mu}\partial_{\mu}\lambda~~~,
\end{equation}
and comparison with the third line then shows that $\psi_{\nu}$ obeys the constraint
\begin{equation}\label{constr}
0=\gamma^{\nu}\psi_{\nu}~~~`
\end{equation}

In terms of the left chiral reduction, Eqs. \eqref{lameq} and \eqref{constr} are
\begin{align}\label{leftr}
\Psi_0=&\vec \sigma \cdot \vec \Psi~~~,\cr
\partial_0\ell=&\vec \sigma \cdot \vec \nabla \ell~~~,\cr
\end{align}
while  the second line of Eq. \eqref{covL} yields both the spin-$\frac{3}{2}$ equation of motion
\begin{equation}\label{eqmo}
\partial_0\vec\Psi= \vec \nabla \Psi_0+i\vec \nabla \times \vec\Psi
\end{equation}
and the constraint
\begin{equation}\label{constr1}
\vec \sigma \cdot \vec \nabla \times \vec \Psi=im \ell~~~.
\end{equation}

From the Dirac brackets for the free field coupled model \cite{adler1} multiplied by $i$, we get the equal time anticommutation relations obeyed by $\vec \Psi$ and $\ell$,
\begin{align}\label{anti}
\{\Psi_i(\vec x),\Psi^{\dagger}_j(\vec y)\}_+=&\frac{1}{2}\sigma_j\sigma_i
\delta^3(\vec x-\vec y)-\vec{\nabla}_{xi} \frac{\delta^3(\vec x-\vec y)}{m^2} \overleftarrow{\nabla}_{yj} \cr
=&( \delta_{ij}-\frac{1}{2}\sigma_i\sigma_j)
\delta^3(\vec x-\vec y)-\vec{\nabla}_{xi} \frac{\delta^3(\vec x-\vec y)}{m^2} \overleftarrow{\nabla}_{yj} ~~~,\cr
\{\ell(\vec x),\Psi^{\dagger}_j(\vec y)\}_+=&-\frac{1}{m}\delta^3(\vec x-\vec y)\overleftarrow{\nabla}_{yj}~~~,\cr
\{\Psi_i(\vec x),\ell^{\dagger}(\vec y)\}_+=&-\frac{1}{m}\vec{\nabla}_{xi}\delta^3(\vec x-\vec y)~~~,\cr
\{\ell(\vec x),\ell^{\dagger}(\vec y)\}=&0~~~.\cr
\end{align}

\subsection{Momentum space analysis}

Let us now write the field equations in momentum space.  For the covariant equations,
we write
\begin{align}\label{momeq1}
\psi_{\mu}(x)=&\frac{1}{(2\pi)^4} \int d^4k e^{ik \cdot x} \psi_{\mu}[k]~~~,\cr
\lambda(x)=&\frac{1}{(2\pi)^4} \int d^4k e^{ik \cdot x} \lambda[k]~~~,\cr
e^{ik\cdot x}=&e^{i(\vec k \cdot \vec x - k^0 x^0)}~~~,\cr
\end{align}
giving in momentum space
\begin{align}\label{momeq2}
-\gamma^{\mu\nu\rho}ik_{\nu}\psi_\rho[k]=&m\gamma^{\mu}\lambda[k]~~~,\cr
\gamma^{\nu}k_{\nu}\lambda[k]=&0~~~,\cr
\gamma^{\nu}\psi_{\nu}[k]=&0~~~.\cr
\end{align}
Similarly, for the left chiral equations we write
\begin{align}\label{momeq3}
\Psi_i(x)=&\frac{1}{(2\pi)^4} \int d^4k e^{ik \cdot x}\Psi_i[k]~~~,\cr
\Psi_0(x)=&\frac{1}{(2\pi)^4} \int d^4k e^{ik \cdot x}\Psi_0[k]~~~,\cr
\ell(x)=&\frac{1}{(2\pi)^4} \int d^4k e^{ik \cdot x}\ell[k]~~~,\cr
\end{align}
giving in momentum space (with $\Omega=k^0$)
\begin{align}\label{momeq4}
\Psi_0[k]=&\vec \sigma \cdot \vec \Psi[k]~~~,\cr
0=&(\Omega + \vec \sigma \cdot \vec k)\ell[k]~~~,\cr
-\Omega \vec \Psi[k]=&\vec k \Psi_0[k]+i\vec k \times \vec \Psi[k]~~~,\cr
0=&\vec \sigma \cdot \vec k \times \vec \Psi[k]-m \ell[k]~~~.\cr
\end{align}
In \cite{adler1} we wrote Eqs. \eqref{momeq4} in terms of a wave operator
$[W]$ acting on $\vec \Psi$ and $\ell$ and found the Jordan eigenmodes, which
are given in Table I (with $\Omega$ reversed in sign from \cite{adler1}).  In this
table, we have defined $K,\, \hat k_+,\, \hat k_-,\,\chi_\uparrow,\,\chi_\downarrow$ by
\begin{align}\label{tabledefs}
K=&|\vec k|~~~,\cr
\hat k\times \hat k_+=&-i \hat k_+~~~,\cr
\hat k\times \hat k_-=&i \hat k_-~~~,\cr
\vec \sigma\cdot \hat k\chi_\uparrow=&\chi_\uparrow~~~,\cr
\vec \sigma\cdot \hat k\chi_\downarrow=&-\chi_\downarrow~~~.\cr
\end{align}

\begin{table} [ht]
\caption{The 6 plane wave modes.}
\centering
\begin{tabular}{ c c c c c c  }
\hline\hline
{\rm eigenvector} & $\vec \Psi[k]$ & ~~~~$\ell[k]$ &~~ helicity & ~~~~$[W]\times {\rm eigenvector}$ &$ ~~~{\rm eigenvalue}=\frac{\Omega}{K}=\frac{k^0}{|\vec k|}$ \\
$v_1$ &$\hat k \chi_{\uparrow}$ &~~ 0&$\frac{1}{2}$ & $v_1$ & $-1$ \\
$v_2$ &$\hat k \chi_{\downarrow}$ &~~ 0&$-\frac{1}{2}$ &  $-v_2$& $1$ \\
$v_3$ &$\hat k_+\chi_{\uparrow}$  &~~ 0&$\frac{3}{2}$ & $v_3$&$-1$ \\
$v_4$ &$\hat k_-\chi_{\downarrow}$ &~~ 0&$-\frac{3}{2}$&$-v_4$ & $1$ \\
$v_5$ &$\frac{1}{2}\hat k_+\chi_{\downarrow}$ & $~~-i\frac{K}{m}\chi_{\uparrow}$  &$\frac{1}{2}$ &$v_5+v_1$&$-1$\\
$v_6$ &$\frac{1}{2}\hat k_-\chi_{\uparrow}$  & $~~~~~i\frac{K}{m}\chi_{\downarrow}$ &$-\frac{1}{2}$ &$-v_6+v_2$& $1$\\
\hline
\hline\hline
\end{tabular}
\label{modes}
\end{table}

\subsection{Mode expansion and equal time anticommutator algebra}

In order to study equal time anticommutation relations, it is convenient to rewrite
Eq. \eqref{momeq3} as
\begin{align}\label{momeq5}
\Psi_i(\vec x,t)=&\frac{1}{(2\pi)^3} \int d^3k e^{i\vec k \cdot \vec x}\Psi_i[\vec k,t]~~~,\cr
\Psi_0(\vec x,t)=&\frac{1}{(2\pi)^3} \int d^3k e^{i\vec k \cdot \vec x}\Psi_0[\vec k,t]~~~,\cr
\ell(\vec x,t)=&\frac{1}{(2\pi)^3} \int d^3k e^{i\vec k \cdot \vec x}\ell[\vec k,t]~~~,\cr
\end{align}
where, for example,
\begin{equation}\label{momeq6}
\Psi_i[\vec k,t]=\frac{1}{2\pi} \int dk^0e^{-ik^0t }\Psi_i[k]~~~,
\end{equation}
and similarly for the second and third lines of Eq. \eqref{momeq5}.  We will also
need the adjoints of $\Psi_i$ and $\ell$, which are given by
\begin{align}\label{momeq7}
\Psi_i^{\dagger}(\vec x,t)=&\frac{1}{(2\pi)^3} \int d^3k e^{-i\vec k \cdot \vec x}\Psi_i^{\dagger}[\vec k,t]~~~,\cr
\ell^{\dagger}(\vec x,t)=&\frac{1}{(2\pi)^3} \int d^3k e^{-i\vec k \cdot \vec x}\ell^{\dagger}[\vec k,t]~~~.\cr
\end{align}

We now introduce creation and annihilation operators $b_i[\vec k,t]~,~b_i^{\dagger}[\vec k, t]~,~~i=1,...,6$ as coefficients in the expansions of $\Psi_i[\vec k, t]$, $\ell[\vec k, t]$ and their adjoints over the basis of Table I, as follows:
\begin{align}\label{expansion}
\vec \Psi[\vec k,t]=&b_1[\vec k,t] \hat k \chi_\uparrow +b_2[\vec k,t] \hat k \chi_\downarrow+b_3[\vec k,t] \hat k_+ \chi_\uparrow + b_4[\vec k,t] \hat k_-\chi_\downarrow + b_5[\vec k,t] \frac{1}{2} \hat k_+ \chi_\downarrow + b_6[\vec k,t]
\frac{1}{2} \hat k_- \chi_\uparrow~~~,\cr
\vec \Psi^{\dagger}[\vec k,t]=&b_1^{\dagger}[\vec k,t] \hat k \chi_\uparrow^{\dagger} +b_2^{\dagger}[\vec k,t] \hat k \chi_\downarrow^{\dagger}+b_3^{\dagger}[\vec k,t] \hat k_- \chi_\uparrow^{\dagger} + b_4^{\dagger}[\vec k,t] \hat k_+\chi_\downarrow^{\dagger} + b_5^{\dagger}[\vec k,t] \frac{1}{2} \hat k_- \chi_\downarrow^{\dagger} + b_6^{\dagger}[\vec k,t]
\frac{1}{2} \hat k_+ \chi_\uparrow^{\dagger}~~~,\cr
\ell[\vec k,t]=&b_5[\vec k,t]\frac{-iK}{m} \chi_\uparrow + b_6[\vec k,t] \frac{iK}{m} \chi_\downarrow~~~,\cr
\ell^{\dagger}[\vec k,t]=&b_5^{\dagger}[\vec k,t]\frac{iK}{m} \chi_\uparrow^{\dagger} + b_6^{\dagger}[\vec k,t] \frac{-iK}{m} \chi_\downarrow^{\dagger}~~~.\cr
\end{align}

To rewrite the canonical anticommutation relations of Eq. \eqref{anti} in terms of
the mode creation and annihilation operators $b_i[\vec k,t]~,~b_i^{\dagger}[\vec k, t]$
we need the following expansion of $\sigma_j\sigma_i$ on a $\hat k,~\hat k_+,~\hat k_-$
basis,
\begin{equation}\label{expansion1}
\sigma_j\sigma_i=\hat k_j\hat k_i - \hat k_j \hat k_{+i} \chi_\downarrow \chi^{\dagger}_\uparrow+\hat k_j\hat k_{-i}\chi_\uparrow \chi^{\dagger}_\downarrow
+\hat k_{+j} \hat k_i \chi_\downarrow \chi^{\dagger}_\uparrow -\hat k_{-j}\hat k_i \chi_\uparrow \chi^{\dagger}_\downarrow+\hat k_{+j} \hat k_{-i} \chi_\downarrow
\chi^{\dagger}_\downarrow+ \hat k_ {-j}\hat k_ {+i} \chi_\uparrow \chi^{\dagger}_\uparrow~~~,
\end{equation}
while the unit Pauli spin operator has the expansion
\begin{equation}\label{expansion2}
1=\chi_\uparrow \chi^{\dagger}_\uparrow + \chi_\downarrow \chi^{\dagger}_\downarrow~~~.
\end{equation}
Substituting the Fourier representation of the Dirac delta function
\begin{equation}\label{delta}
\delta^3(\vec x-\vec y)= \frac{1}{(2\pi)^3} \int d^3 k e^{i \vec k \cdot (\vec x-\vec y)}
\end{equation}
into the right-hand side of Eq. \eqref{anti}, and substituting the expansions of
Eqs. \eqref{momeq5}, \eqref{momeq7}, \eqref{expansion1}, and \eqref{expansion2}  into the left-hand side of Eq. \eqref{anti}, we get the following results:  The anticommuators
are given by
\begin{equation}\label{anti1}
\{b_i(\vec k,t),b_j^{\dagger}(\vec k^{\prime},t)\}_+=(2\pi)^3 \delta^3(\vec k- k^{\prime})  C_{ij}~~~,
\end{equation}
with $C_{ij}$ a numerical matrix the entries of which are zero except for the nonzero
elements shown in Tables II, III, and IV, with the matrix element $a$ given by
\begin{equation}\label{adef}
a=\frac{1}{2}-\frac{K^2}{m^2}~~~.
\end{equation}

\begin{table} [ht]
\caption{Nonzero anticommutators in the 3--4 sector}
\centering
\begin{tabular}{c c c c}
\hline\hline
~~~~~&\vline&$b_3^\dagger$  &  $b_4^\dagger$\\
\hline
$b_3$ & \vline&$\frac{1}{2}$  &  $0$ \\
$b_4$ &\vline& $0$  &   $\frac{1}{2}$\\
\hline\hline
\end{tabular}
\label{threefour}
\end{table}

\begin{table} [ht]
\caption{Nonzero anticommutators in the 1--5 sector.  The body of the table
defines an anticommutator matrix $M_{III}$}
\centering
\begin{tabular}{c c c c}
\hline\hline
~~~~~&\vline&$b_1^\dagger$  &  $b_5^\dagger$\\
\hline
$b_1$ & \vline& $a$  &  $-1$ \\
$b_5$ &\vline& $-1$  &   $0$\\
\hline\hline
\end{tabular}
\label{onefive}
\end{table}

\begin{table} [ht]
\caption{Nonzero anticommutators in the 2--6 sector.  The body of the table
defines an anticommutator matrix $M_{IV}$}
\centering
\begin{tabular}{c c c c}
\hline\hline
~~~~~&\vline&$b_2^\dagger$  &  $b_6^\dagger$\\
\hline
$b_2$ & \vline&$a$  &  $1$ \\
$b_6$ &\vline& $1$  &   $0$\\
\hline\hline
\end{tabular}
\label{twosix}
\end{table}

 We see that the off-diagonal entries in the $1-5$ and $2-6$ sectors correspond
 to the Jordan blocks that appear in Table I.

 \subsection{Indefinite metric structure}

 As noted in \cite{adler1}, the anticommutators of Eq. \eqref{anti} require an indefinite metric Hilbert space.  To show the indefinite metric structure explicitly, we
 diagonalize the anticommutator matrices $M_{III}$ and $M_{IV}$ given in Tables III and IV.  The characteristic
 equation for both is
 \begin{equation}\label{char1}
 {\rm det} (M_{III}-\lambda 1) = {\rm det} (M_{IV}-\lambda 1)= \lambda^2-\lambda a -1=0~~~,
 \end{equation}
with roots given by
\begin{align}\label{roots}
\lambda_{\pm}=&a/2 \pm (1+a^2/4)^{1/2}~~~,\cr
\lambda_+ &>0~~~,~~~\lambda_- <0~~~,\cr
\lambda_+\lambda_- =&-1~~~,\cr
\lambda_+ + \lambda_- = &a~~~.\cr
\end{align}
We now define diagonalized operators in the $2-6$ and $1-5$ sectors by
\begin{align}\label{diags}
D_{\pm}=&\lambda_{\pm} b_2 +b_6~~~,\cr
E_{\pm}=&\lambda_{\pm} b_1 -b_5~~~,\cr
\end{align}
which obey
\begin{align}\label{diags1}
\{D_+,D_-^\dagger\}_+=&\{E_+,E_-^\dagger\}_+=0~~~,\cr
\{D_+,D_+^\dagger\}_+=&\{E_+,E_+^\dagger\}_+=\lambda_+ F_+~~~,\cr
\{D_-,D_-^\dagger\}_+=&\{E_-,E_-^\dagger\}_+=\lambda_- F_-~~~.\cr
\end{align}
Here $F_{\pm}$ are given by
\begin{equation}\label{fdef}
F_{\pm}=2+a \lambda_{\pm}~~~,
\end{equation}
and obey $F_+F_-=F_++F_-=4+a^2>0$, which implies that both $F_+$ and $F_-$ are
positive.  Hence the anticommutators $\{D_+,D_+^\dagger\}_+$ and $\{E_+,E_+^\dagger\}_+$ are both
positive, and the anticommutators $\{D_-,D_-^\dagger\}_+$ and $\{E_-,E_-^\dagger\}_+$ are both
negative.  Thus the construction of this section localizes the indefinite metric in the $D_-,~E_-$
sector of Hilbert space.

\section{Propagator Structure}

\subsection{Left chiral propagator}

In \cite{adler1} we constructed the manifestly covariant propagator derived from the Lagrangian density of Eq. \eqref{covL}. In this section we construct the corresponding
left chiral propagator.  We start from the left chiral action given in Eq. (9) of
\cite{adler1}, which rewritten in momentum space takes the form (again with $\Omega=k^0$)
\begin{align}\label{chiralS}
S=&\frac{1}{(2\pi)^4}\int d^4k S[k]~~~,\cr
S[k]=&-i\Psi_0^\dagger[k] (\vec \sigma \times \vec k) \cdot \vec \Psi[k] +
i \vec{\Psi}^\dagger[k] \cdot (\vec \sigma \times \vec k) \Psi_0[k] + i \vec{\Psi}^{\dagger}[k]\cdot \vec k \times \vec{\Psi}[k] + i \Omega \vec{\Psi}^{\dagger}[k] \cdot
\vec \sigma \times \vec{\Psi}[k]\cr
 +& \Omega \ell^\dagger[k]\ell[k] +\ell^\dagger[k] \vec \sigma
\cdot \vec k \ell[k] +im(-\ell^{\dagger}[k] \Psi_0[k] + \ell^\dagger[k] \vec \sigma \cdot \vec{\Psi}[k]
+ \Psi_0^\dagger[k] \ell[k] - {\vec \Psi}^{\dagger}[k] \cdot \vec \sigma \ell[k])~~~\cr
=&\Big(\Psi_i^\dagger[k] ~ \Psi_0^\dagger[k] ~ \ell^\dagger[k] \Big)\cal{\tilde M}  \left( \begin{array} {c}
 \Psi_j[k]  \\ \Psi_0[k] \\
 \ell[k] \\  \end{array}\right)~~~\cr
\end{align}
with $\cal{\tilde M}$ the matrix
\begin{equation}\label{mdef}
\cal{\tilde M}=\left( \begin{array} {c c c}
 i \epsilon_{ipj}(k_p+\Omega \sigma_p)    & i(\vec \sigma \times \vec k)_i
  & -im  \sigma_i    \\
 -i (\vec \sigma \times \vec k)_j   &   0    &  im      \\
im \sigma_j     &  -im     & \Omega+\vec \sigma \cdot \vec k  \end{array}\right)   ~~~.   \\
\end{equation}
The propagator for the coupled fields is the matrix $\cal{\tilde N}$ that is inverse to
$\cal{\tilde M}$,
\begin{align}\label{ndef}
{\cal \tilde M} {\cal \tilde N}=& \left( \begin{array} {c c c}
\delta_{i\ell}  &  0  & 0   \\
 0  & 1   &  0  \\
 0  & 0   &  1  \\
\end{array}\right)~~~,\cr
{\cal \tilde N}=&\left( \begin{array} {c c c}
\tilde N_{1j\ell}  &  \tilde N_{2j}  & \tilde N_{3j}   \\
\tilde N_{4\ell} & \tilde N_5   &  \tilde N_6  \\
\tilde N_{7\ell}  & \tilde N_8   &  \tilde N_9  \\
\end{array}\right)~~~.\cr
\end{align}
Recalling that $\overline{\psi}_\mu=\psi_\mu^\dagger i \gamma^0
\,,~\overline{\ell}=\ell^\dagger i \gamma^0$, the simplest way to calculate ${\cal \tilde N}$ is by left chiral projection from the covariant form given Eqs. (67)-(70) of \cite{adler1},  by making the substitutions
\begin{align}\label{subs}
\overline{\psi}_{\mu} \to & i({\Psi}_i^\dagger,{\Psi}_0^{\dagger})P_R~~~,\cr
\overline{\lambda} \to & i\ell^{\dagger} P_R~~~,\cr
\psi_{\rho} \to & P_L\left( \begin{array} {c}
\Psi_j  \\ \Psi_0  \\
 \end{array}\right)~~~,\cr
\lambda \to  & P_L \ell ~~~,\cr
 \end{align}
with $P_R=\frac{1}{2}(1-\gamma_5)$ and $P_L=\frac{1}{2}(1+\gamma_5)$ the right
and left chiral projectors.  Defining
\begin{align}\label{qdef}
Q\equiv & 2(\frac{1}{m^2}-\frac{2}{k^2})~~~,\cr
a_{\pm}\equiv& \vec k \cdot \vec \sigma \pm \Omega~~~,\cr
\end{align}
and remembering that $k^2=(\vec k)^2-(k^0)^2=K^2-\Omega^2$, we get
\begin{align}\label{Nresults}
\tilde N_{1j\ell}=&\frac{-1}{2k^2}(\sigma_\ell a_+ \sigma_j + Q k_j k_\ell a_-)~~~,\cr
\tilde N_{2j}=&\frac{1}{2k^2}(a_+ \sigma_j+Q k_j \Omega a_-)~~~,\cr
\tilde N_{3j}=&\frac{-i}{mk^2} k_j a_-~~~,\cr
\tilde N_{4\ell}=&\frac{1}{2k^2}(\sigma_{\ell} a_+ + Q  \Omega k_\ell a_-)~~~,\cr
\tilde N_5=&\frac{-1}{2k^2}(a_+ + Q \Omega^2 a_-)~~~,\cr
\tilde N_6=&\frac{i}{mk^2} \Omega a_-~~~,\cr
\tilde N_{7\ell}=&\frac{i}{mk^2} k_\ell a_-~~~,\cr
\tilde N_8=&\frac{-i}{mk^2} \Omega a_-~~~,\cr
\tilde N_9=&0~~~.\cr
\end{align}
Substituting these into $\cal{\tilde N}$ of Eq. \eqref{ndef}, and using Eq. \eqref{mdef}, one
can verify that $\cal \tilde N$ is the inverse of $\cal \tilde M$. Related to the fact that
$\Psi_0 =\vec \sigma \cdot \vec \Psi$, we find that the entries in Eq. \eqref{Nresults}
obey the constraint equations
\begin{align}\label{constraintrels}
\tilde N_{1j\ell}\sigma_\ell=&\tilde N_{2j}-k_j/m^2~~~,\cr
\sigma_j \tilde N_{1j\ell}=&\tilde N_{4\ell}-k_\ell/m^2~~~,\cr
\sigma_j\tilde N_{2j}=&\tilde N_5+\Omega/m^2~~~,\cr
\tilde N_{4\ell} \sigma_{\ell}=&\tilde N_5+\Omega/m^2~~~,\cr
\sigma_j\tilde N_{3j}=&\tilde N_6-i/m~~~,\cr
\tilde N_{7\ell}\sigma_{\ell}=&\tilde N_8+i/m~~~.\cr
\end{align}
The corresponding constraint relations obeyed by the entries $N_{1\rho\sigma}$ and
$N_{2\rho}$ of the covariant propagator $\cal N$ of Eq. (69) of \cite{adler1} are
\begin{align}\label{covconstraint}
\gamma^{\rho}N_{1\rho\sigma}=&-ik_\sigma/m^2~~~,\cr
N_{1\rho\sigma}\gamma^{\sigma}=&-ik_\rho/m^2~~~,\cr
\gamma^{\rho}N_{2\rho}=&1/m~~~,\cr
N_{3\sigma}\gamma^{\sigma}=&-1/m~~~.\cr
\end{align}
The presence of inhomogeneous terms in Eqs. \eqref{constraintrels} and \eqref{covconstraint} is a reflection of the fact that the action  inversions to
obtain the covariant and left chiral propagators are done before enforcing
the constraint $\Psi_0 =\vec \sigma \cdot \vec \Psi$; this constraint is a consequence
of the equations of motion following from the action but is not substituted back
into the action.  When $m\to \infty$, from the Lagrangian in Eq. \eqref{covL} we see that the constraint
is enforced off shell (i.e., without use of the equations of motion) and the inhomogeneous terms then
all vanish.

\subsection{Normalization constant relating $\cal N$ to the Fourier transform of
the time ordered product}

According to standard quantum field theory, we expect the left chiral propagator
$\cal N$ to be the Fourier transform of the time ordered product of fields, up to
a normalization constant that we now  determine.  That is, we expect to find relations
\begin{align}\label{tprod}
\langle 0|T\big(\Psi_j(x)\Psi_\ell^\dagger(y)\big)|0\rangle=&
\int \frac{d\Omega}{2\pi}\frac{d^3 k}{(2\pi)^3}e^{i(\vec k \cdot \vec x-\Omega x^0)}
c \tilde N_{1j\ell}~~~,\cr
\langle 0|T\big(\Psi_j(x)\ell^\dagger(y)\big)|0\rangle=&
\int \frac{d\Omega}{2\pi}\frac{d^3 k}{(2\pi)^3}e^{i(\vec k \cdot \vec x-\Omega x^0)}
c \tilde N_{3j}~~~,\cr
\langle 0|T\big(\ell(x)\Psi_\ell^\dagger(y)\big)|0\rangle=&
\int \frac{d\Omega}{2\pi}\frac{d^3 k}{(2\pi)^3}e^{i(\vec k \cdot \vec x-\Omega x^0)}
c \tilde N_{7\ell}~~~,\cr
\langle 0|T\big(\ell(x)\ell^\dagger(y)\big)|0\rangle=&
\int \frac{d\Omega}{2\pi}\frac{d^3 k}{(2\pi)^3}e^{i(\vec k \cdot \vec x-\Omega x^0)}
c \tilde N_{9}=0~~~.\cr
\end{align}
We will elaborate on this correspondence in subsequent sections, where we
use it to determine the action of the creation and annihilation operators on the
vacuum state $|0\rangle$. But first we must check the consistency of Eq. \eqref{tprod} with the anticommutation relations of Eq. \eqref{anti}, and use this to determine
the constant $c$.

To do this we use the equation of motion of Eq. \eqref{eqmo}, which after elimination of $\Psi_0$ by use of the constraint on the first line of Eq. \eqref{leftr} can be written as
\begin{align}\label{eqmo1}
0=&(\partial_0 \delta_{jp}-M_{jp})\Psi_p(x)~~~,\cr
M_{jp}=&\partial_j \sigma_p+i\epsilon_{jmp}\partial_m~~~.\cr
\end{align}
Applying the differential operator $\partial_0\delta_{jp}-M_{jp}$ to the first line on the left-hand
side of Eq. \eqref{tprod}, we get
\begin{align}\label{diffop1}
(\partial_0\delta_{jp}-M_{jp})\langle 0|T\big(\Psi_p(x)\Psi_\ell^\dagger(0)\big)\rangle=& \delta(x^0) \langle 0|\{\Psi_j(x),\Psi_\ell^\dagger(0)\}_+|0\rangle\cr
=&\delta(x^0) \left(\frac{1}{2}\sigma_\ell\sigma_j+\frac{\partial_j\partial_\ell}{m^2}\right)
\delta^3(\vec x) \cr
=&\int \frac{d\Omega}{2\pi}\frac{d^3 k}{(2\pi)^3}e^{i(\vec k \cdot \vec x-\Omega x^0)}
\left(\frac{1}{2}\sigma_\ell\sigma_j-\frac{k_jk_\ell}{m^2}\right)~~~.\cr
\end{align}
So consistency with Eq. \eqref{tprod} requires
\begin{equation}\label{consist1}
i(-\Omega \delta_{jp}-k_j\sigma_p-i\epsilon_{jmp}k_m) c \tilde  N_{1p\ell}=\frac{1}{2}\sigma_\ell\sigma_j-\frac{k_jk_\ell}{m^2}~~~.
\end{equation}
By some algebra starting from the first line of Eq. \eqref{Nresults}, one can verify that the identity of
Eq. \eqref{consist1} is obeyed with $c=i$.  Similarly, one can verify that with this choice of $c$ the
remaining lines  of Eq. \eqref{tprod} are consistent with the equations of motion for $\vec \Psi$ and $\ell$
and the equal time anticommutators of Eq. \eqref{anti}.  Because of the inhomogeneous terms in Eq. \eqref{constraintrels} analogous
relations do not hold for $T$ products of $\Psi_0,\, \Psi_0^\dagger$ with the other fields.

\subsection{Propagator construction from action of creation and annihilation operators
on the vacuum}

Focusing now on the first line of Eq. \eqref{tprod}, let us examine how the propagator is constructed from free particle creation
and annihilation operators acting on the vacuum state.  Expanding the time ordered product, the left-hand side of the first line  (specialized
with no loss of generality to $y=0$) is
\begin{equation}\label{leftside}
\langle 0|T\big(\Psi_j(x)\Psi_\ell^\dagger(0)\big)|0\rangle=\theta(x^0)\langle 0|\Psi_j(x)\Psi_\ell^\dagger(0)|0\rangle
-\theta(-x^0)\langle 0|\Psi_\ell^\dagger(0)\Psi_j(x)|0\rangle~~~,
\end{equation}
with $\theta(x^0)$ the Heaviside step function.  Taking $c=i$, the right-hand side of the first line with $y=0$ is
\begin{equation}\label{rightside}
\int \frac{d\Omega}{2\pi}\frac{d^3 k}{(2\pi)^3}e^{i(\vec k \cdot \vec x-\Omega x^0)}
\frac{-i}{2(k^2-i\epsilon)}(\sigma_{\ell}a_+\sigma_j+Qk_jk_\ell a_-)~~~,
\end{equation}
with $k^2=\vec k^2-(k^0)^2$, $a_{\pm}=\vec k\cdot \vec \sigma \pm k^0$, and $Q=2\big(1/m-2/(k^2-i\epsilon)\big)$ from Eq. \eqref{qdef}.
In the denominators we have introduced the usual Feynman $i\epsilon$ prescription to give the rule for moving the poles at $k^2=0$ off
the real $\Omega=k^0$ integration axis, to $k^0=K-i\epsilon$ and $k^0=-K+i\epsilon$, with $K=|\vec k|$.   This integration has the unusual feature of having a double pole in $k^2-i\epsilon$, which is not seen
in the usual textbook propagators but is also encountered in quantum electrodynamics in covariant gauges  \cite{lautrup}.  Carrying out the $k^0$
integration by closing the contour up for $x^0<0$ and down for $x^0>0$, Eq. \eqref{rightside} becomes
\begin{equation}\label{evaluatedrhs}
\int \frac{d^3 k}{(2\pi)^3}\left[e^{i(\vec k \cdot \vec x-K x^0)}\frac{1}{2}\theta(x^0)\Lambda_{j\ell}^+
+ e^{i(\vec k \cdot \vec x+K x^0)}\frac{1}{2}\theta(-x^0)\Lambda_{j\ell}^-\right]~~~,
\end{equation}
with
\begin{align}\label{pluspart}
\Lambda_{j\ell}^+ =&\sigma_{\ell} \frac{\vec \sigma \cdot \vec k+K}{2K} \sigma_j+ \frac{2k_jk_\ell}{m^2}\frac{\vec \sigma \cdot \vec k-K}{2K}
-\frac{k_jk_\ell}{K^2}\left[ix^0(\vec \sigma \cdot \vec k-K)+\frac{\vec \sigma \cdot \vec k}{K}\right]~~~,\cr
\Lambda_{j\ell}^- =&\sigma_{\ell} \frac{\vec \sigma \cdot \vec k-K}{2K} \sigma_j+ \frac{2k_jk_\ell}{m^2}\frac{\vec \sigma \cdot \vec k+K}{2K}
+\frac{k_jk_\ell}{K^2}\left[ix^0(\vec \sigma \cdot \vec k+K)-\frac{\vec \sigma \cdot \vec k}{K}\right]~~~.\cr
\end{align}
Comparing Eq. \eqref{leftside} with Eq. \eqref{pluspart}, we see that we need to show that the action of the creation and annihilation
operators on the vacuum implies that
\begin{align}\label{separated}
\langle 0|\Psi_j(x)\Psi_\ell^\dagger(0)|0\rangle=&\int \frac{d^3 k}{(2\pi)^3}e^{i(\vec k \cdot \vec x-K x^0)}\frac{1}{2}\Lambda_{j\ell}^+~,~~x^0\geq 0~~~,\cr
-\langle 0|\Psi_\ell^\dagger(0)\Psi_j(x)|0\rangle=&\int \frac{d^3 k}{(2\pi)^3}e^{i(\vec k \cdot \vec x+K x^0)}\frac{1}{2}\Lambda_{j\ell}^-~,~~x^0\leq 0~~~.\cr
\end{align}
We shall do this in two stages, by first considering the equal time match at $x^0=0$, and then extending to
consider the cases $x^0>0\,,~x^0<0$ by using the Heisenberg representation time evolution of $\Psi_j(x)$.

\subsection{The equal time match}

Specializing to $x^0=0$, the matching requirement of Eqs. \eqref{separated} and \eqref{pluspart} becomes
\begin{align}\label{separated1}
\langle 0|\Psi_j(\vec x,0)\Psi_\ell^\dagger(0)|0\rangle=&\int \frac{d^3 k}{(2\pi)^3}e^{i\vec k \cdot \vec x}\frac{1}{2}(D_{j\ell}+C_{j\ell})~~~,\cr
-\langle 0|\Psi_\ell^\dagger(0)\Psi_j(\vec x,0)|0\rangle=&\int \frac{d^3 k}{(2\pi)^3}e^{i\vec k \cdot \vec x}\frac{1}{2}(D_{j\ell}-C_{j\ell})~~~,\cr
\end{align}
where we have written $\Lambda_{j\ell}^\pm=D_{j\ell}\pm C_{j\ell}$, with
\begin{align}\label{dcdefs}
D_{j\ell}=&\frac{1}{2}\sigma_{\ell}\vec \sigma \cdot \hat k \sigma_j+\vec \sigma \cdot \hat k \hat k_j
\hat k_\ell \left(\frac{K^2}{m^2}-1\right)~~~,\cr
C_{j\ell}=&\frac{1}{2}\sigma_\ell\sigma_j-\frac{k_jk_\ell}{m^2}~~~.\cr
\end{align}
The difference of the two lines in Eq. \eqref{separated1} is the Fourier transformed anticommutator match
\begin{equation}\label{diffmatch}
\langle 0|\Psi_j(\vec x,0)\Psi_\ell^\dagger(0)+\Psi_\ell^\dagger(0)\Psi_j(\vec x,0)|0\rangle
=\int \frac{d^3 k}{(2\pi)^3}e^{i\vec k \cdot \vec x}C_{j\ell}~~~,
\end{equation}
which is already guaranteed by the nonzero anticommutators of Tables II-IV.  So we only have
to check the sum of the two lines in Eq. \eqref{separated1},
\begin{equation}\label{summatch}
\langle 0|\Psi_j(\vec x,0)\Psi_\ell^\dagger(0)-\Psi_\ell^\dagger(0)\Psi_j(\vec x,0)|0\rangle
=\int \frac{d^3 k}{(2\pi)^3}e^{i\vec k \cdot \vec x}D_{j\ell}~~~.
\end{equation}
Expanding $\sigma_{\ell}\vec \sigma \cdot \hat k \sigma_j$ on a $\hat k,~\hat k_+,~\hat k_-$
basis, we get in analogy with Eq. \eqref{expansion1}
\begin{align}\label{expansion3}
\sigma_{\ell}\vec \sigma \cdot \hat k \sigma_j=&\hat k_\ell \hat k_j(\chi_\uparrow \chi^{\dagger}_\uparrow-\chi_\downarrow \chi^\dagger_\downarrow)+\hat k_\ell \hat k_{+j} \chi_\downarrow \chi^\dagger_\uparrow +\hat k_\ell \hat k_{-j} \chi_\uparrow \chi^\dagger_\downarrow +\hat k_{+\ell}\hat k_j
\chi_\downarrow \chi^\dagger_\uparrow +\hat k_{-\ell}\hat k_j \chi_\uparrow \chi^\dagger_\downarrow \cr
+&\hat k_{+\ell}\hat k_{-j} \chi_\downarrow \chi^\dagger_\downarrow - \hat k_{-\ell}\hat k_{+j}\chi_\uparrow \chi^\dagger_\uparrow~~~,\cr
\end{align}
which when substituted into Eq. \eqref{dcdefs} and remembering that $a=1/2-K^2/m^2$ gives the analogous
expansion for $D_{j\ell}$,
\begin{align}\label{expansion4}
D_{j\ell}=&-a\hat k_\ell \hat k_j(\chi_\uparrow \chi^{\dagger}_\uparrow-\chi_\downarrow \chi^\dagger_\downarrow)+\frac{1}{2}[\hat k_\ell \hat k_{+j} \chi_\downarrow \chi^\dagger_\uparrow +\hat k_\ell \hat k_{-j} \chi_\uparrow \chi^\dagger_\downarrow +\hat k_{+\ell}\hat k_j
\chi_\downarrow \chi^\dagger_\uparrow +\hat k_{-\ell}\hat k_j \chi_\uparrow \chi^\dagger_\downarrow \cr
+&\hat k_{+\ell}\hat k_{-j} \chi_\downarrow \chi^\dagger_\downarrow - \hat k_{-\ell}\hat k_{+j}\chi_\uparrow \chi^\dagger_\uparrow]~~~.\cr
\end{align}
This puts the right-hand side of Eq. \eqref{summatch} in a form where it can be compared with the left-hand
side after substitution of the Fourier representations of Eq. \eqref{momeq5} and \eqref{momeq6} together with  the mode expansions of Eq. \eqref{expansion}.  We find that the mode anticommutators of Tables II, III, and IV imply that all terms match if we adopt the following Dirac hole interpretation of the mode creation and annihilation operators:  For the modes 2, 4, 6 with $\Omega/K=1$, the corresponding $b$ operator is the  annihilation operator, while for the modes 1, 3, 5 with $\Omega/K=-1$, the corresponding $b^{\dagger}$ operator is the annihilation operator.  That is, acting on the vacuum state we require
\begin{align}\label{vacuumaction}
b_j|0\rangle =&0~~,~~~\langle 0|b_j^{\dagger}=0~~,~~~j=2,4,6~~~,\cr
b_j^\dagger |0\rangle =&0~~,~~~\langle 0|b_j=0~~,~~~j=1,3,5~~~\cr
\end{align}
in order to be able to construct the propagator from creation and annihilation operator action on the
vacuum state.

\subsection{The Hamiltonian and time evolution of creation and annihilation operators}

To do the unequal time match, we will use the Heisenberg representation time evolution formula
$\vec \Psi(\vec x, t)= e^{iHt} \vec \Psi(\vec x, 0)e^{-iHt}$, with $H$ the coupled model Hamiltonian
\begin{equation}\label{ham1}
H=\int d^3 x [ -\vec \Psi^\dagger \cdot (\vec \nabla \times \vec \Psi) +i \ell^\dagger \vec \sigma \cdot \vec \nabla \ell + im(\vec \Psi^\dagger \cdot \vec \sigma \ell - \ell^\dagger \vec \sigma \cdot \vec \Psi)]~~~.
\end{equation}
Expressed in terms of the Fourier transforms of the fields, this is
\begin{equation}\label{ham2}
H=\frac{1}{(2\pi)^3} \int d^3k [-\vec \Psi^\dagger[\vec k,t] \cdot i\vec k \times \vec \Psi[\vec k,t]
 -\ell^\dagger[\vec k,t] \vec \sigma \cdot \vec k \ell[\vec k,t]+im(\vec \Psi^\dagger[\vec k,t]\cdot \vec \sigma \ell[\vec k,t]-\ell^\dagger[\vec k,t]\vec \sigma \cdot \vec \Psi[\vec k,t])]~~~.
\end{equation}
Substituting the mode expansions of Eq. \eqref{expansion}, and normal ordering to put annihilation
operators on the right, this becomes (with arguments $[\vec k,t]$
suppressed)
\begin{equation}\label{ham3}
H=\frac{1}{(2\pi)^3} \int d^3k K [2(b_4^\dagger b_4+b_3 b_3^\dagger) -(a+1) (b_5 b_5^\dagger + b_6^\dagger b_6) - (b_5 b_1^\dagger + b_1 b_5^\dagger) + b_2^\dagger b_6+ b_6^\dagger b_2)]~~~.
\end{equation}
This can be written as the sum of three mutually commuting pieces, $H=H_{34}+H_{15}+H_{26}$, with
\begin{align}\label{ham4}
H_{34}=&\frac{1}{(2\pi)^3} \int d^3k K 2(b_4^\dagger b_4+b_3 b_3^\dagger)~~~,\cr
H_{15}=&\frac{1}{(2\pi)^3} \int d^3k K [ -(a+1) b_5 b_5^\dagger - (b_5 b_1^\dagger +b_1b_5^\dagger)]~~~,\cr
H_{26}=&\frac{1}{(2\pi)^3} \int d^3k K [ -(a+1) b_6^\dagger b_6 + b_2^\dagger b_6+ b_6^\dagger b_2)]~~~.\cr
\end{align}
Since the Hamiltonian is time independent, we are free to take the time argument $t$  in Eqs. \eqref{ham3}
and \eqref{ham4} as $t=0$, which we shall do in the remainder of this section.

We now use these three separate pieces of $H$ to  determine the time evolution of the mode operators in the corresponding sectors given in Tables II, III, and IV.  Beginning with the $3-4$ sector in Table II,
we define
\begin{align}\label{f34def}
f_3(\vec k, t)=&e^{iHt}b_3(\vec k, 0)e^{-iHt}~~~,\cr
f_4(\vec k, t)=&e^{iHt}b_4(\vec k, 0)e^{-iHt}~~~.\cr
\end{align}
Differentiating with respect to time and using the anticommutators of Table II we get
\begin{align}\label{f34diff}
(d/dt) f_3(\vec k,t)=& ie^{iHt}[H_{34},b_3(\vec k, 0)]e^{-iHt}=iKf_3(\vec k,t)~~~,\cr
(d/dt) f_4(\vec k,t)=& ie^{iHt}[H_{34},b_4(\vec k, 0)]e^{-iHt}=-iKf_4(\vec k,t)~~~,\cr
\end{align}
which can be immediately integrated to give
\begin{align}\label{f34ev}
f_3(\vec k,t)=& e^{iKt} f_3(\vec k,0)~~~,\cr
f_4(\vec k,t)=& e^{-iKt} f_4(\vec k,0)~~~,\cr
\end{align}
that is,
\begin{align}\label{f34ev1}
e^{iHt}b_3(\vec k, 0)e^{-iHt}=&e^{iKt}b_3(\vec k,0)~~~,\cr
e^{iHt}b_4(\vec k, 0)e^{-iHt}=&e^{-iKt}b_4(\vec k,0)~~~.\cr
\end{align}

Turning next to the $1-5$ sector in Table III, we define
\begin{align}\label{f15def}
f_1(\vec k, t)=&e^{iHt}b_1(\vec k, 0)e^{-iHt}~~~,\cr
f_5(\vec k, t)=&e^{iHt}b_5(\vec k, 0)e^{-iHt}~~~.\cr
\end{align}
Differentiating with respect to time we get
\begin{equation}\label{f15diff}
(d/dt) f_{1,5}(\vec k,t)= ie^{iHt}[H_{15},b_{1,5}(\vec k, 0)]e^{-iHt}~~~,
\end{equation}
which using the anticommutators of Table III gives the coupled differential equations
\begin{align}\label{f15diff1}
(d/dt)f_1(\vec k, t)=& iK[f_1(\vec k, t)+f_5(\vec k, t)]~~~,\cr
(d/dt)f_5(\vec k, t)=& iKf_5(\vec k, t)~~~.\cr
\end{align}
The second of these equations can be immediately integrated to give
\begin{equation}\label{f5ev}
f_5(\vec k,t)=e^{iKt} f_5(\vec k,0)~~~;
\end{equation}
substituting this into the first equation and integrating then gives
\begin{equation}\label{f1ev}
f_1(\vec k,t)=e^{iKt}[f_1(\vec k,0)+iKt f_5(\vec k,0)]~~~,
\end{equation}
that is
\begin{align}\label{f15ev1}
e^{iHt} b_1(\vec k,0)e^{-iHt}=& e^{iKt}[b_1(\vec k,0)+iKt b_5(\vec k,0)]~~~,\cr
e^{iHt} b_5(\vec k,0)e^{-iHt}=& e^{iKt} b_5(\vec k,0)~~~.\cr
\end{align}
An analogous calculation for the $2-6$ sector, using the anticommutators of Table IV, gives
\begin{align}\label{f26ev1}
e^{iHt} b_2(\vec k,0)e^{-iHt}=& e^{-iKt}[b_2(\vec k,0)+iKt b_6(\vec k,0)]~~~,\cr
e^{iHt} b_6(\vec k,0)e^{-iHt}=& e^{-iKt} b_6(\vec k,0)~~~.\cr
\end{align}

For the application of the next section, we only need the time evolution of the operators $b_{1,...,6}$,
but by taking the adjoint of Eqs. \eqref{f34ev1}, \eqref{f15ev1} and \eqref{f26ev1} we can immediately get
the time evolution of their adjoints $b_{1,...,6}^\dagger$.

\subsection{The $t\neq 0$ match}
Returning to Eq. \eqref{separated}, we now do the $t\neq 0$ match, by substituting the time evolution formulas of Eqs. \eqref{f34ev1},
  \eqref{f15ev1}, and  \eqref{f26ev1} into the mode expansion for $\vec \Psi[\vec k,t]$ of Eq. \eqref{expansion}, and then substituting this into Eq. \eqref{separated}.
Two types of $t$ dependence are present, 
exponential factors $e^{\mp i K t}$ multiplying $\Lambda_{j\ell}^{\pm}$, and additional explicit factors of $t$  in $\Lambda_{j\ell}^{\pm}$.
For terms that do not have an explicit factor of $t$, the exponentials match because $\langle 0|b_{1,3,5}=0$ and $b_{2,4,6}|0\rangle=0$, guaranteeing that the terms in $\Psi_j[\vec k,t]$ with the wrong sign exponential do not contribute.  The pieces in $e^{\mp iKt}\frac{1}{2}\Lambda_{j\ell}^{\pm}$ with an explicit factor of $t$ are
\begin{equation}\label{explicit}
\mp e^{\mp i K t} \frac{1}{2} iKt \hat k_j \hat k_{\ell} (\hat k \cdot \vec \sigma \mp 1)=e^{\mp i K t} i K t \hat k_j \hat k_{\ell}  \left( \begin{array} {c}
 \chi_\downarrow \chi^{\dagger}_\downarrow  \\ \chi_\uparrow \chi^{\dagger}_\uparrow\\
  \end{array}\right)~~~.
\end{equation}
In $\Psi_j(x)$ on the first line of Eq. \eqref{separated} only the explicit $t$ in front of $b_6$  contributes, giving
\begin{equation}\label{pluscase}
\langle 0|e^{-iKt} iKt b_6 \hat k_j \chi_\downarrow b_2^{ \dagger} \hat k_{\ell} \chi_{\downarrow}^\dagger
|0\rangle~~~,
\end{equation}
which  matches, while in
 $\Psi_j(x)$ on the second line of Eq. \eqref{separated} only the explicit $t$ in front of $b_5$  contributes, giving
\begin{equation}\label{minuscase}
-\langle 0| b_1^\dagger \hat k_{\ell}  e^{iKt} iKt  b_5  \hat k_j  \chi_{\uparrow}\chi_\uparrow^\dagger
|0\rangle~~~,
\end{equation}
which again matches (where in ordering the spinor factors we have used the fact that the second line in Eq. \eqref{separated} is a spinor outer product, not a 
spinor inner product).  This completes the demonstration that the propagator pieces corresponding to the first line of Eq. \eqref{tprod} are correctly constructed from the left-hand side
of Eq. \eqref{separated} using the algebra of the creation and annihilation operators.

\subsection{Propagator construction for the second line of Eq. \eqref{tprod}}

With $c=i$, the second line of Eq. \eqref{tprod} is
\begin{align}\label{tprod1}
\langle 0|T\big(\Psi_j(x)\ell^\dagger(y)\big)|0\rangle=&
\int \frac{d\Omega}{2\pi}\frac{d^3 k}{(2\pi)^3}e^{i(\vec k \cdot \vec x-\Omega x^0)}
i \tilde N_{3j}\cr
=&\int \frac{d\Omega}{2\pi}\frac{d^3 k}{(2\pi)^3}e^{i(\vec k \cdot \vec x-\Omega x^0)}
\frac{k_j}{mk^2}(\vec k \cdot \vec \sigma -\Omega)~~~.\cr
\end{align}
Carrying out the $\Omega=k^0$ integration with the Feynman $i\epsilon$ prescription, we find for the
right-hand side of Eq. \eqref{tprod1}
\begin{equation}\label{tprod2}
\int \frac{d^3 k}{(2\pi)^3}e^{i\vec k \cdot \vec x} \frac{iK}{2m}\hat k_j[\theta(x^0)e^{-iKx^0}(\vec \sigma \cdot \hat k -1) + \theta(-x^0) e^{iKx^0} (\vec \sigma \cdot \hat k +1)]~~~.
\end{equation}
Separating into positive and negative $x^0$ pieces, the left-hand side of Eq. \eqref{tprod} becomes
\begin{equation}\label{seppsiell}
\theta(x^0)\langle 0| \Psi_j(x) \ell^{\dagger}(0)|0\rangle - \theta(-x^0) \langle 0| \ell^{\dagger}(0) \Psi_j(x)  |0\rangle~~~.
\end{equation}
Using the vacuum state conditions of Eq. \eqref{vacuumaction} and the anticommuators of Tables III and IV,
we find that Eqs. \eqref{tprod2} and \eqref{seppsiell} match.  The propagator construction and matching
conditions for the third line of Eq. \eqref{tprod} can be similarly verified.  The match for the fourth
line of Eq. \eqref{tprod} is an immediate consequence of the mode expansion of Eq. \eqref{expansion} for $\ell$  and $\ell^\dagger$, together with the zero anticommutators on the lower right diagonals of Tables III and IV.

\section{Conclusion}

We have given a detailed analysis of the free field structure of the model \cite{adler1} in which a
spin-$\frac{3}{2}$ Rarita-Schwinger field is directly coupled to a spin-$\frac{1}{2}$ field.  We have shown that many properties, such as the action of the creation and annihilation
operators on the vacuum state, the Dirac hole construction, and the propagator construction using the Feynman $i\epsilon$ prescription, are the ones familiar from conventional spin-$\frac{1}{2}$ theories.  Other features are novel. One is the indefinite metric sector in Hilbert space, which we have related to the diagonalization of the anticommutators of Tables III and IV, which in turn follow form the Jordan
eigenmode structure of the wave operator.  Another is the propagator double pole which gives rise to an explicit  $t=x^0$ term in the time ordered products.  The next step in studying the coupled model is to include
the gauge field interactions, as formulated in \cite{adler1}, and to see if these can trigger spontaneous
 symmetry breaking, so that the fields of the model develop Dirac masses and the left- and right- chiral
sectors of the model no longer decouple, or the fields develop Majorana masses within the left-chiral sector.

\section{Acknowledgements}

Completion of this work was supported
in part by the National Science Foundation under Grant No. PHYS-1066293 through  the hospitality of the Aspen Center for Physics.

\end{document}